\documentclass[11pt,draftcls,onecolumn]{IEEEtran}
\usepackage{graphicx}
\usepackage{fancyhdr}
\usepackage{amsmath,amssymb,amstext}
\usepackage{slashbox}
\usepackage{bm}
\usepackage{algorithm}
\usepackage{algorithmic}
\usepackage{array}
\usepackage{cite}
\usepackage{subfigure}
\usepackage{color}

\newtheorem{proposition}{\textbf{Proposition}}

\begin{document}
%
\title{Reconfigurable Intelligent Surface With Energy Harvesting Assisted Cooperative Ambient Backscatter Communications}
%
%

\author{Hui~Ma, 
        Haijun~Zhang, \IEEEmembership{Senior~Member,~IEEE,}
        Ning~Zhang, \IEEEmembership{Senior~Member,~IEEE,}
        Jianquan~Wang,
        Ning~Wang, \IEEEmembership{Member,~IEEE,}
        ~and~Victor~C.~M.~Leung, \IEEEmembership{Life~Fellow, ~IEEE}
        \thanks{Hui Ma, Haijun Zhang and Jianquan Wang are with University of Science and Technology Beijing, Beijing 100083, China (e-mail: hui\_ma@ustb.edu.cn, haijunzhang@ieee.org, wangjianquan@ustb.edu.cn).}
        \thanks{Ning Zhang is with the Department of Electrical and Computer Engineering, University of Windsor, Windsor, ON N9B 3P4, Canada (e-mail: Ning.Zhang@uwindsor.ca).}
        \thanks{Ning Wang is with Henan Joint International Research Laboratory of Intelligent Networking and Data Analysis, School of Information Engineering, Zhengzhou University, Zhengzhou 450000, China (email: ienwang@zzu.edu.cn).}
        \thanks{Victor C. M. Leung is with the Department of Electrical and Computer	Engineering, University of British Columbia, Vancouver, BC V6T 1Z4, Canada (e-mail: vleung@ieee.org).}
}

%



\maketitle

\begin{abstract}

The performance of cooperative ambient backscatter communications (CABC) can be enhanced by employing reconfigurable intelligent surface (RIS) to assist backscatter transmitters. Since the RIS power consumption is a non-negligible issue, we consider a RIS assisted CABC system where the RIS with energy harvesting circuit can not only reflect signal but also harvest wireless energy. We study a transmission design problem to minimize the RIS power consumption with the quality of service  constraints for both active and backscatter transmissions. The optimization problem is a mixed-integer non-convex programming  problem which is NP-hard. To tackle it, an algorithm is proposed by employing the block coordinate descent, semidefinite relaxation and alternating direction method of multipliers techniques. Simulation results demonstrate the effectiveness of the proposed algorithm. 

\end{abstract}


%
\IEEEpeerreviewmaketitle

\section{Introduction}
\label{sect:introduction}

Cooperative ambient backscatter communication (CABC) is a promising solution for low-energy
Internet of Things. In CABC, a backscatter transmitter (B-Tx) can send information to a cooperative receiver (C-Rx) by modulating and reflecting ambient radio frequency (RF) signals without employing active components, such as oscillators and power amplifiers. The B-Tx realizes cooperative ambient backscatter transmission such that the C-Rx can jointly decode the messages from it and the active transmitter (A-Tx) which transmits RF signals. Thus, the cooperative ambient backscatter transmission can not only deliver information but also improve the active transmission in return by providing additional multipath. {\textcolor{black}{In conventional CABC, B-Txs are equipped with single reflecting antenna \cite{Yang2018,Zhou2019}. However, due to the double fading effect, the backscatter links are weak. The performance of the backscatter transmissions and the enhancement to the active transmissions are limited.}}

{\textcolor{black}{To address the aforementioned issue, reconfigurable intelligent surface (RIS) can be employed in CABC to assist B-Txs \cite{Zhang2021}. A RIS is a planar surface consisting of a controller and  a large number of elements. These elements can reflect signals and induce phase shifts to the reflected signals, through which backscatter transmission can be performed based on phase shift keying (PSK). Moreover, by appropriately setting phase shifts at RIS elements, fine-grained reflect beamforming can be achieved \cite{CSIBeam}. Through reflect beamforming, RIS has great potential to improve the backscatter and active transmissions.}}

{\textcolor{black}{In conventional CABC, the power consumption on single reflecting antenna is assumed negligible. Similarly, existing research on RIS assisted CABC seldom considered the power consumption on RIS \cite{Zhang2021}. However, for a low-energy B-Tx which powers a RIS, the RIS power consumption  is not negligible, since the number of reflecting elements is large \cite{Huang2019}. This motivates us to study RIS assisted CABC, taking RIS power consumption into account.}} In this letter, we consider a  RIS assisted CABC system. The RIS is equipped with energy harvesting (EH) circuit such that each RIS element can operate under either signal reflecting mode or EH mode. With the help of EH elements, the RIS power consumption can be reduced. We minimize the RIS power consumption under quality of service (QoS) constraints for both active and backscatter transmissions by assigning the modes of RIS elements and optimizing the RIS reflect beamforming and the transmit beamforming at the multi-antenna A-Tx. The transmission design problem is a mixed-integer non-convex programming (MINCP) problem which is NP-hard \cite{Zhanghaijun2020}. To tackle the MINCP problem, we develop an algorithm by employing the block coordinate descent (BCD), semidefinite relaxation (SDR) and alternating direction method of multipliers (ADMM) techniques. We reveal that a rank-one optimal solution can always be constructed for the SDR problem. In the ADMM iteration procedure, closed form solutions are given for the optimization problems occurred. Simulation results are provided that demonstrate the effectiveness of the proposed algorithm.
\vspace{-0.2cm}
\section{System Model and Problem Formulation}
\label{sect:system:model}
\begin{figure}
\begin{center}
\includegraphics[width=0.9\linewidth, draft=false]{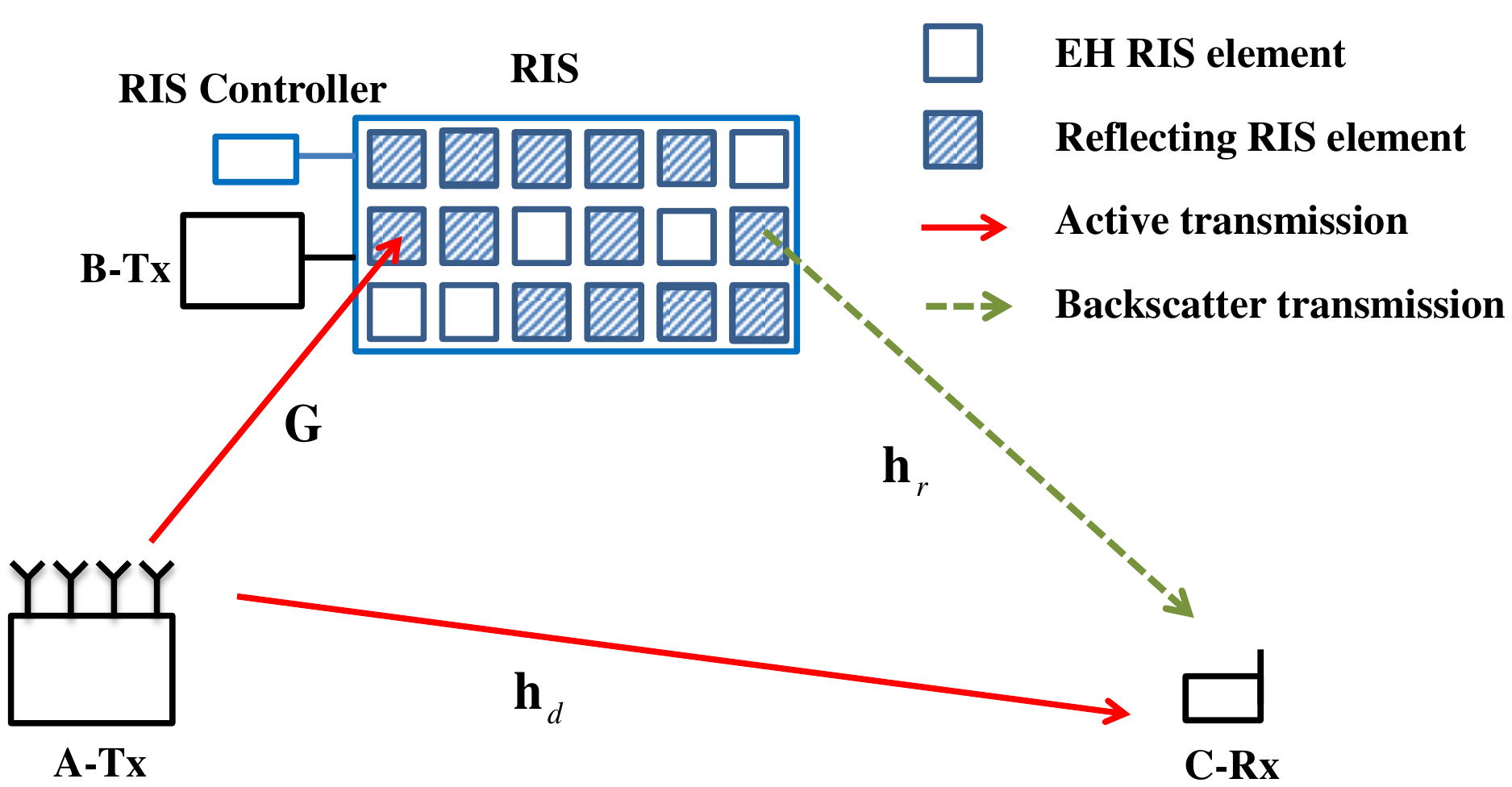}
\caption{A RIS assisted CABC system.}
\vspace{-0.4cm}
\label{FigModel}
\end{center}
\end{figure}
We consider a CABC system composed of an A-Tx with $N$ antennas, a low-energy B-Tx assisted by a RIS, and a single-antenna C-Rx. The B-Tx and the RIS are connected by a wire and the RIS consists of $I_R$ RIS elements, denoted by a set ${\mathcal{I}} = \{ 1,...,I_R\}$.
The A-Tx transmits signals to the C-Rx via transmit beamforming. The transmit beamforming vector is denoted by ${\bf{w}} \in  {\mathbb{C}}^{N \times 1}$. {\textcolor{black}{In the meanwhile, the B-Tx sends messages to the C-Rx by utilizing RIS to implement backscatter modulation, i.e., modulating its symbols over the ambient radio frequency signals from the A-Tx through periodically changing the phase shift at each reflecting element.}} In the backscatter transmission, the RIS reflects incident signals via reflect beamforming and binary PSK is applied. Thus, the $i$-th RIS element uses $c\psi_i$ for reflecting signals, where $c\in\mathcal{C}=$ $\{1,-1\}$ denotes the transmitted symbol at the B-Tx, $\psi_i=e^{j\phi_i}$ is the reflect beamforming parameter at the $i$-th RIS element and  $\phi_i \in \mathcal{A}= [0,2\pi)$ is the beamforming phase shift. 
{\textcolor{black}{The RIS is equipped with EH circuit such that each RIS element can operate under either signal reflecting mode or EH mode \cite{EH}.}} 
Let $s_i, \forall i \in {\mathcal{I}}$ denote the mode assignment of the $i$-th RIS element: {\textcolor{black}{when $s_i=0$, RIS element $i$ is assigned to operate under EH mode; when  $s_i=1$,  RIS element $i$ is assigned to operate under reflecting mode.}}
Then, the reflection coefficient at the $i$-th RIS element can be expressed as  $cs_i\psi_i$. The period of backscatter transmission symbol $c$ covers $L(L \gg 1)$ periods of active transmission symbols. Let $x(1),...,x(L)$ denote the active transmission symbols covered by $c$ with  ${\mathbb{E}}[|x(l)|^2]=1,  l ={1,...,L}$. Then, the received signal at C-Rx can be expressed as
\begin{equation}
\label{RxSig}
y(l) = {\bf{h}}_d^H{\bf{w}}x(l) + \sqrt \alpha  {\bf{h}}_r^Hc{{\bf{S}}\bm{\Psi}} {\bf{Gw}}x(l) + n(l)
\end{equation}
for $l= 1,...,L$, where $\alpha$ denotes the reflection efficiency and $n(l)$ is the noise at the C-Rx that follows distribution ${\mathcal{CN}}(0,\sigma^2)$ and ${\mathcal{CN}}(0,\sigma^2)$ denotes the circularly symmetric complex Gaussian
distribution with zero mean and variance $\sigma^2$. ${\bf{h}}_d \in {\mathbb{C}}^{N \times 1}$, ${\bf{h}}_r \in {\mathbb{C}}^{I_R \times 1}$ and ${\bf{G}} \in {\mathbb{C}}^{I_R \times N}$ represent the complex channels from A-Tx to C-Rx, from {\textcolor{black}{RIS}} to C-Rx and from A-Tx to {\textcolor{black}{RIS}}, respectively. All the channels are block flat fading channels. ${\bf{S}}=\text{diag}({s_1,...,s_{I_R}})$ and ${\bm{\Psi}}=\text{diag}({\psi_1,...,\psi_{I_R}})$ are the RIS element mode assignment diagonal matrix and the reflect beamforming diagonal matrix, where $\text{diag}(\bullet)$ is the diagonal matrix construction operator. The C-Rx jointly decode $x(l)$ and $c$ from the received signal. According to \eqref{RxSig}, the signal-to-noise-ratio (SNR) for decoding $x(l)$ can be written as \cite{Zhang2021}
\begin{equation}
\Gamma_A = \frac{\left( {{\bf{h}}_d^H + c\sqrt \alpha  {\bf{h}}_r^H{\bf{S}}{\bm{\Psi}}{\bf{G}}} \right){\bf{w}}{\bf{w}}^H\left( {{\bf{h}}_d^{} + c\sqrt \alpha  {{\bf{G}}^H}{\bm{\Psi}}^H{\bf{S}}^H{\bf{h}}_r^{}} \right)}{{{\sigma ^2}}} \nonumber
\end{equation}
and the SNR for decoding $c$ can be given by 
\begin{equation}
\Gamma_B= \frac{{\alpha L}}{{{\sigma ^2}}}{\bf{h}}_r^H{\bf{S}}{\bm{\Psi}}{\bf{G}}{\bf{w}}{\bf{w}}^H{{\bf{G}}^H}{\bm{\Psi}}^H{\bf{S}}^H{\bf{h}}_r^{}.  \nonumber
\end{equation}

{\textcolor{black}{The RIS is powered by the B-Tx. The power consumption on RIS is mainly caused by  controlling the phase shift at each reflecting element \cite{Zhu,Lyu,ZouS}. On the other hand, EH elements can reduce the RIS power consumption by collecting wireless energy. By denoting the power consumption on controlling each reflecting element's phase shifting as $u$, the RIS power consumption can be expressed as}}
\begin{equation}
E_{\text{RIS}}({{\bf{w}},{{\bf{S}}}})={\rm{tr}}({\bf{S}})u - \eta {\rm{tr}}\left( {\left( {{\bf{I}} - {\bf{S}}} \right){\bf{Gw}}{{\bf{w}}^H}{{\bf{G}}^H}\left( {{\bf{I}} - {\bf{S}}} \right)} \right) \nonumber
\end{equation}
where $\eta$ denotes the EH efficiency\footnote{{\textcolor{black}{In this letter, we consider the linear energy harvesting model, the more complicated cases will be studied in the future works.}}}.

Our goal is to minimize the RIS power consumption\footnote{{\textcolor{black}{Besides the power consumed by RIS, the other power dissipation at the B-Tx is constant during the backscatter transmission. Therefore we only focus on minimizing the RIS power consumption. 
}}} for the B-Tx while ensuring the QoSs of the active and the backsactter transmissions by assigning the modes of RIS elements and optimizing the transmit beamforming and the reflect beamforming. Based on the discussion above, the transmission design problem can be formulated as
\begin{equation}
\begin{split}
\nonumber
({\bf{P}}) :  \hspace{0.2cm} &\mathop {\min }\limits_{{\bf{w}},{\rm{ }}{\bf{S}},{\rm{ }}{\bm{\Psi}}} {\rm{ }} E_{\text{RIS}}({{\bf{w}},{{\bf{S}}}}) \\
&s.t. \hspace{0.1cm} {\rm{C}}1: \Gamma_A \ge {\gamma _A}, \forall c \in \mathcal{C}, {\rm{ }} {\rm{C}}2: \Gamma_B \ge {\gamma _B}, \\
&\hspace{0.6cm} {\rm{C}}3: {\rm{tr}}({\bf{w}}{{\bf{w}}^H}) \le P_{bgt},   {\rm{ }} {\rm{C}}4: |\psi_i|=1, \forall i \in \mathcal{I},\\
&\hspace{0.6cm} {\rm{C}}5: {s_i} = \{ 0,1\}, \forall i \in \mathcal{I}\\
\end{split}
\end{equation}
where $\gamma_A$ and $\gamma_B$ are the SNR requirements of the active and the backscatter transmissions, respectively, and $P_{bgt}$ is the transmit power budget of the A-Tx. 
In problem (${\bf{P}}$), constraints C1 and C2 ensure the QoS. Constraint C3 guarantees that the transmit power at the A-Tx does not exceed its transmit power budget. Constraint C4 is imposed because the reflect beamforming parameter $\psi_i$ takes the form $\psi_i=e^{j\phi_i}$, $\phi_i \in \mathcal{A}$. Constraint C5 specifies that each RIS element can only operate under either reflecting mode or EH mode.
\vspace{-0.2cm}
\section{Algorithm Development}
\label{sect:Algorithm}

The optimization problem (${\bf{P}}$) is a MINCP problem since the optimization variables ${\bf{w}}$ and ${\bm{\Psi}}$ are coupled in constraints C1 and C2, and $s_i$ is a binary optimization variable. If the binary optimization variables are tackled by the exhaustive search method, the search space is $\mathcal{O}(2^{I_R})$. Since the number of RIS elements, i.e, $I_R$ is large, applying the exhaustive search leads to extremely high computation complexity. In addition, even though the exhaustive search is adopted, it is still hard to find an optimal solution to  (${\bf{P}}$) because of the coupling optimization variables. In this section, we propose an efficient algorithm based on the BCD technique to find a suboptimal solution to problem  (${\bf{P}}$), where the optimization variables are decoupled into two blocks and are optimized alternatively by leveraging the SDR and ADMM techniques.

\vspace{0.1cm}
\noindent\textit{A. Transmit Beamforming Optimization}

According to the original problem  (${\bf{P}}$), with fixed RIS element mode assignment diagonal matrix  ${\bf{S}}$ and reflect beamforming diagonal matrix ${\bm{\Psi}}$, the transmit beamforming vector ${\bf{w}}$ can be optimized through solving the following optimization problem
\begin{equation}
\nonumber
({\bf{P1.1}}) :  \hspace{0.2cm} \mathop {\min }\limits_{{\bf{w}}} {\rm{ }}E_{\text{RIS}}({{\bf{w}},{{\bf{S}}}}), \hspace{0.1cm} s.t. \hspace{0.1cm} {\rm{C}}1, {\rm{ }} {\rm{C}}2, {\rm{ }} {\rm{C}}3.
\end{equation}
However, problem (${\bf{P1.1}}$) is not convex due to the non-convex constraints C1 and C2. To make problem (${\bf{P1.1}}$) tractable, we define ${\bf{W}}={\bf{w}}{\bf{w}}^H$, which implies $\rm{rank}({\bf{W}})=1$. By ignoring the rank-one constraint on ${\bf{W}}$, the SDR of problem (${\bf{P1.1}}$) can be represented as
\begin{equation}
\begin{split}
\nonumber
&({\bf{P1.2}}) : \\
&\mathop {\min }\limits_{\bf{W}} {\rm{ tr}}({\bf{S}})u - \eta {\rm{tr}}\left( {\left( {{\bf{I}} - {\bf{S}}} \right){\bf{GW}}{{\bf{G}}^H}\left( {{\bf{I}} - {\bf{S}}} \right)} \right)\\
&s.t. \hspace{0.1cm} \frac{\left( {{\bf{h}}_d^H + \sqrt \alpha  {\bf{h}}_r^H{\bf{S\Psi G}}} \right){\bf{W}}\left( {{\bf{h}}_d^{} + \sqrt \alpha  {{\bf{G}}^H}{{\bf{\Psi }}^H}{\bf{S}}^H{\bf{h}}_r^{}} \right)}{{{\sigma ^2}}} \ge {\gamma _A}, \\
&\hspace{0.6cm}\frac{\left( {{\bf{h}}_d^H - \sqrt \alpha  {\bf{h}}_r^H{\bf{S\Psi G}}} \right){\bf{W}}\left( {{\bf{h}}_d^{} - \sqrt \alpha  {{\bf{G}}^H}{{\bf{\Psi }}^H}{\bf{S}}^H{\bf{h}}_r^{}} \right)}{{{\sigma ^2}}} \ge {\gamma _A},\\
&\hspace{0.6cm} \frac{{\alpha L}}{{{\sigma ^2}}}{\bf{h}}_r^H{\bf{S\Psi GW}}{{\bf{G}}^H}{{\bf{\Psi }}^H}{{\bf{S}}^H}{\bf{h}}_r^{} \ge {\gamma _B},\\
&\hspace{0.6cm} {\rm{tr}}({\bf{W}}) \le P_{bgt},{\bf{W}}\succeq {\bf{0}}.
\end{split}
\end{equation} 
Problem  (${\bf{P1.2}}$) is a convex semidefinite programming (SDP) problem and an optimal solution to it can be derived by the interior-point method. Nevertheless, since it is a relaxed version, solving it does not necessarily guarantee that an optimal solution to (${\bf{P1.1}}$) can be found. In order to solve  (${\bf{P1.1}}$), we have the following proposition. 

\begin{proposition}\label{p:solution}
In  problem ({\bf{P1.2}}), a rank-one optimal solution can always be constructed, through which an optimal solution to  ({\bf{P1.1}}) can be obtained, as shown in Algorithm 1. 
\end{proposition}

\emph{Proof}: The proof can be obtained by following the one in \cite[Appendix A]{Huang2010}.
\begin{algorithm}[h] \label{AL1}
	\caption{Transmit beamforming optimization algorithm}
{\small	\begin{algorithmic}[1]
		\STATE Solve the convex SDP problem ({\bf{P1.2}}) to find an optimal solution $\bf{W}^*$ to it.
		
		\STATE If {$\rm{rank}({\bf{W}}^*)>1$}, go to step 3. Otherwise, set ${{\bf{W}}^+}={{\bf{W}}^*}$ and go to step 7.
		
		\STATE Decompose ${{\bf{W}}^*} = {\bf{V}}{{\bf{V}}^H}$.
		
		\STATE	Find a non-zero Hermitian matrix ${\bf{Z}}$ that satisfies the equations ${\rm{tr}}\left( {{{\bf{V}}^H}{{\bf{G}}^H}{{\bf{\Psi }}^H}{\bf{S}}^H{\bf{h}}_r^{} {\bf{h}}_r^H{\bf{S\Psi G}}{\bf{VZ}}} \right)= 0$, ${{\rm{tr}}\left( {{{\bf{V}}^H}{\bf{h}}_d^{}{\bf{h}}_d^H{\bf{VZ}}} \right) = 0}$ and  ${{\rm{tr}}\left( {{{\bf{V}}^H}{\bf{VZ}}} \right) = 0}$.

		\STATE If the eigenvalues of  ${\bf{Z}}$ are not all less than 0, compute ${{\bf{W}}^+} = {\bf{V}}({\bf{I}} - (1/{\rm{maxeig(}}{\bf{Z}})){\bf{Z}}){{\bf{V}}^H}$; otherwise, compute ${{\bf{W}}^+} = {\bf{V}}({\bf{I}} - (1/{\rm{mineig(}}{\bf{Z}})){\bf{Z}}){{\bf{V}}^H}$, where $\rm{maxeig}({\bf{Z}})$ and  $\rm{mineig}({\bf{Z}})$ represent the maximum and minimum eigenvalues of matrix ${\bf{Z}} $ respectively.
		
		\STATE If {$\rm{rank}({\bf{W}}^+)=1$}, go to step 7. Otherwise, set ${{\bf{W}}^*}={{\bf{W}}^+}$ and go to step 3.
		
		\STATE ${{\bf{W}}^+}$ is a rank-one optimal solution to  Problem  (${\bf{P1.2}}$) and ${{\bf{w}}^+} = \sqrt {{\rm{maxeig}}\left( {{{\bf{W}}^+}} \right)} {\rm{maxeigvc}}\left( {{{\bf{W}}^+}} \right)$ is an optimal solution to  problem ({\bf{P1.1}}), i.e, the optimal transmit beamforming vector, where ${\rm{maxeigvc}}\left( {{{\bf{W}}^+}} \right)$ represents the eigenvector corresponding to the maximum eigenvalue of ${{{\bf{W}}^+}}$.
	\end{algorithmic}   }
\end{algorithm}

\noindent\textit{B. RIS Element Mode Assignment and Reflect Beamforming Optimization}

In this subsection, we optimize the RIS element mode assignment diagonal matrix  ${\bf{S}}$ and reflect beamforming diagonal matrix ${\bm{\Psi}}$. With the transmit beamforming vector ${\bf{w}}$ being fixed, the original problem  (${\bf{P}}$) can be simplified into 
 \vspace{-0.2cm}
 \begin{equation}
 \nonumber
 ({\bf{P2.1}}) :  \hspace{0.2cm} \mathop {\min }\limits_{{\bf{S}},{\rm{ }}{\bm{\Psi}}} {\rm{ }} E_{\text{RIS}}({{\bf{w}},{{\bf{S}}}}), \hspace{0.1cm} s.t. \hspace{0.1cm} {\rm{C}}1,{\rm{ }} {\rm{C}}2,{\rm{ }} {\rm{C}}4,{\rm{ }} {\rm{C}}5. 
 \end{equation}
In constraints C1 and C2,  ${\bf{S}}$ and ${\bm{\Psi}}$ are coupled. To decouple the optimization variables, we rewrite ({\bf{P2.1}}) as 
 \begin{equation}
 \begin{split}
 \nonumber
({\bf{P2.2}}):\, &\mathop {\min }\limits_{{\bf{S}},{\rm{ }}{\bm{\theta}}} E_{\text{RIS}}({{\bf{w}},{{\bf{S}}}})\\
 &s.t. \hspace{0.1cm} {\rm{C}}5, \,\,\,\, \frac{1}{{{\sigma ^2}}}{\left| {{\bf{h}}_d^H{\bf{w}} + c\sqrt \alpha  {{\bm{\theta }}^H}{\bf{a}}} \right|^2} \ge {\gamma _A}, \forall c\in \mathcal{C}\\
 &\hspace{0.6cm}\frac{{\alpha L}}{{{\sigma ^2}}}{{\bm{\theta }}^H}{\bf{aa}}_{}^H{\bm{\theta }} \ge {\gamma _B},\,\, {\left| {{\theta _i}} \right|^2} = {s_i},\forall i \in \mathcal{I}
 \end{split}
\end{equation} 
where ${\bf{a}} = {\rm{diag}}({\bf{h}}_r^H){\bf{Gw}}$, ${\bm{\theta }} = {[{\theta _1},...,{\theta _{{I_R}}}]^T}$ and
  ${\theta _i} = {\left( {{s_i}{\psi _i}} \right)^\dag }$ where $(\bullet)^\dag$ is the conjugate operator. By introducing a mapping variable $t$ and auxiliary variables ${\bf{x}}_0$, ${\bf{x}}_1$ and ${\bf{x}}_2$, problem ({\bf{P2.2}}) can be equivalently transformed into  

\begin{equation}
 \begin{split}
 \nonumber
 ({\bf{P2.3}}): &\mathop {\min }\limits_{{\bf{S}},{\rm{ }}{\bm{\bar \theta }},{\rm{ }}{\bf{x}}_0,{\rm{ }}{\bf{x}}_1,{\rm{ }}{\bf{x}}_2}E_{\text{RIS}}({{\bf{w}},{{\bf{S}}}})\\
 &s.t. \hspace{0.1cm}{\rm{C}}5, \,{\rm{C}}6: {\left| {{\bar\theta _i}} \right|^2} = {s_i},\forall i \in \mathcal{I},\, {\rm{C}}7: {\left| {{\bar\theta _{I_R+1}}} \right|^2}  = 1,\\
 &\hspace{0.6cm}  {\rm{C}}8: \frac{1}{{{\sigma ^2}}}{\left| {{\bf{c}}_1^H{{\bf{x}}_1}} \right|^2} \ge {\gamma _A}, {\rm{C}}9: \frac{1}{{{\sigma ^2}}}{\left| {{\bf{c}}_2^H{{\bf{x}}_2}} \right|^2} \ge {\gamma _A},\\
 &\hspace{0.6cm} {\rm{C}}10:\frac{{\alpha L}}{{{\sigma ^2}}}{\left| {{\bf{b}}_{}^H{\bf{x}}_0} \right|^2} \ge {\gamma _B}, \,\,{\bm{\bar \theta }} = {\bf{x}}_0= {{\bf{x}}_1} = {{\bf{x}}_2} 
 \end{split}
 \end{equation}  
where $\bar \theta _i$ represents the $i$-th element in vector ${\bm{\bar \theta }}$ and 
\vspace{0.1cm}

\centerline{${\bm{\bar \theta }} = \left[ {\begin{array}{*{20}{c}}
	{{\bm{\theta }}t}\\
	t
	\end{array}} \right]$,${\bf{b}} = \left[ {\begin{array}{*{20}{c}}
{\bf{a}}\\
0
\end{array}} \right]$,${{\bf{c}}_1} = \left[\begin{array}{*{20}{c}}
{\sqrt \alpha  {\bf{a}}}\\
{{\bf{h}}_d^H{\bf{w}}}
\end{array}\right]$,${{\bf{c}}_2} = \left[\begin{array}{*{20}{c}}
{ - \sqrt \alpha  {\bf{a}}}\\
{{\bf{h}}_d^H{\bf{w}}}
\end{array}\right]$.}\vspace{0.15cm}
In the following, we deal with problem ({\bf{P2.3}}) by using the ADMM technique\footnote{ {\textcolor{black}{Most recently, authors in \cite{VAMP} employed the vector approximate message passing (VAMP) technique for optimizing the reflect beamforming in a RIS assisted communication system. The VAMP technique can be used, because the optimization problem is a mean square
			error minimization problem with only component-wise constraints. Problem ({\bf{P2.3}}) does not take such a form and is more complicated. First, there exist other constraints  besides the component-wise constraints C6 and C7. Second, owing to the optimization variable ${\bf{S}}$, ({\bf{P2.3}}) is an MINCP problem. Therefore, the VAMP technique cannot be applied here.}}  }. The augmented Lagrangian function of  ({\bf{P2.3}}) can be expressed as 
\begin{equation}
L_g=E_{\text{RIS}}({{\bf{w}},{{\bf{S}}}})+\rho {\textstyle{\sum_{m=0}^{2}}}{\left\| {{{\bf{x}}_m} - {\rm{ }}{\bm{\bar \theta }} + {\bm{\mu }}_m} \right\|^2} 
\end{equation}
where ${\bm{\mu }}_0$, ${\bm{\mu }}_1$ and ${\bm{\mu }}_2$ are dual variables associated with  ${{\bf{x}}_0}$, ${{\bf{x}}_1}$ and ${{\bf{x}}_2}$ respectively and $\rho$ is the penalty parameter. In the ADMM procedure,  global variables  ${\bf{S}}$, ${\bm{\bar \theta }}$, auxiliary variables ${{\bf{x}}_0}$, ${{\bf{x}}_1}$, ${{\bf{x}}_2}$ and dual variables ${\bm{\mu }}_0$, ${\bm{\mu }}_1$, ${\bm{\mu }}_2$ are updated iteratively based on the augmented Lagrangian function. Specifically, in the $(n+1)$-th iteration of the ADMM procedure, given the result of the previous iteration, i.e, ${\bf{S}}^{(n)}$,${\bm{\bar \theta }}^{(n)}$ ${\bf{x}}_0^{(n)}$, ${\bf{x}}_1^{(n)}$, ${{\bf{x}}_2^{(n)}}$, ${\bm{\mu }}_0^{(n)}$, ${\bm{\mu }}_1^{(n)}$ and ${\bm{\mu }}_2^{(n)}$, we update the above variables as follows.

(a) Global Variables Update

According to the augmented Lagrangian function, the global variables ${\bf{S}}$, ${\bm{\bar \theta }}$ are updated through 
\begin{equation}
\begin{split}
\nonumber
&\{{\bf{S}}^{(n+1)},{\rm{ }}{\bm{\bar \theta }}^{(n+1)}\} =\\
&\mathop {{\rm{argmin}} }\limits_{{\bf{S}},{\rm{ }}{\bm{\bar \theta }}} E_{\text{RIS}}({{\bf{w}},{{\bf{S}}}})+\rho {\textstyle{\sum_{m=0}^{2}}}{\left\| {{{\bf{x}}_m^{(n)}} - {\rm{ }}{\bm{\bar \theta }} + {\bm{\mu }}_m^{(n)}} \right\|^2} \\
&s.t. \hspace{0.1cm}{\rm{C}}5,\,{\rm{C}}6, \, {\rm{C}}7
\end{split}
\end{equation}  
which can be equivalently rewritten as 
\begin{equation}
\begin{split}
\nonumber
&({\bf{P2.4}}):\, \{{\bf{S}}^{(n+1)},{\rm{ }}{\bm{\bar \theta }}^{(n+1)}\} =\\
&\mathop {{\rm{argmin}} }\limits_{{\bf{S}},{\rm{ }}{\bm{\bar \theta }}} E_{\text{RIS}}({{\bf{w}},{{\bf{S}}}})\hspace{-0.1cm}+\hspace{-0.1cm}{3\rho \sum\limits_{i\in \mathcal{I}} {{s_i}} }-\hspace{-0.1cm}2\rho\Re \left({\bm{\bar \theta }}^H {{\sum_{m=0}^{2}}}{ ({{{\bf{x}}_m^{(n)}} \hspace{-0.05cm} + \hspace{-0.05cm}{\bm{\mu }}_m^{(n)}}) } \right)\\
&s.t. \hspace{0.1cm}{\rm{C}}5,\,{\rm{C}}6, \, {\rm{C}}7.
\end{split}
\end{equation} 

From constraints C$6$ and C$7$,  it can be observed that, for given ${\bf{S}}$, ${\bm{\bar \theta }}^{(n+1)}$ can be obtained by 
\begin{equation}
\label{Obthetai}
\hspace{-0.02cm}\bar \theta _i^{(n+1)} = {s_i}\exp \left( j\arg \left( {\textstyle{{\sum_{m=0}^{2}}}} ( x_{m,i}^{(n)} + \mu_{m,i}^{(n)} ) \right) \right), \forall i \in \mathcal{I},\hspace{-0.2cm}
\end{equation}
\begin{equation}
\label{Obt}
\bar \theta _{I_R+1}^{(n+1)}= \exp \left( j\arg \left( {\textstyle{{\sum_{m=0}^{2}}}} ( x_{m,I_R+1}^{(n)} + \mu_{m,I_R+1}^{(n)} ) \right) \right)
\end{equation}
where $x_{0,i}^{(n)}$, $x_{1,i}^{(n)}$, $x_{2,i}^{(n)}$, $\mu_{0,i}^{(n)}$, $\mu_{1,i}^{(n)}$ and $\mu_{2,i}^{(n)}$ represent the $i$-th elements in vectors ${\bf{x}}_0^{(n)}$, ${\bf{x}}_1^{(n)}$, ${{\bf{x}}_2^{(n)}}$, ${\bm{\mu }}_0^{(n)}$, ${\bm{\mu }}_1^{(n)}$ and ${\bm{\mu }}_2^{(n)}$ respectively. By taking \eqref{Obthetai} and \eqref{Obt} into problem ({\bf{P2.4}}), we have 
\begin{equation}
\label{Obtsi}
s_i^{(n+1)} = \left\{ {\begin{array}{*{20}{l}}
	{0, \,\,\,\, \text{if}\,\,J_i^{(n+1)} > 0}\\
	{1, \,\,\,\, \text{otherwise}}
	\end{array}} \right. \,\,\forall i \in \mathcal{I}
\end{equation}
for deriving ${\bf{S}}^{(n+1)}$, where $J_i^{(n+1)}=\eta {\left| {{g_i}} \right|^2} + u + 3\rho  - 2\rho \left|{\textstyle{{\sum_{m=0}^{2}}}} ( x_{m,i}^{(n)} + \mu_{m,i}^{(n)} ) \right|$ and $g_i$ is the $i$-th element in vector ${\bf{g}}=\bf{Gw}$.
Thus, through \eqref{Obthetai}-\eqref{Obtsi}, ${\bf{S}}^{(n+1)}$ and ${\bm{\bar \theta }}^{(n+1)}$ can be directly obtained.

(b) Auxiliary Variables Update

The auxiliary variables ${{\bf{x}}_0}$, ${{\bf{x}}_1}$ and ${{\bf{x}}_2}$ are updated via the following optimization problem
\begin{equation}
\begin{split}
	\nonumber
	&({\bf{P2.5}}):  \, \{ {\bf{x}}_{0}^{(n+1)},{\bf{x}}_1^{(n+1)},{\bf{x}}_2^{(n+1)} \} =\\
	&\mathop {{\rm{argmin}} }\limits_{{\bf{x}}_{0},{\bf{x}}_1,{\bf{x}}_2} 
	{\mathop{\hspace{-0.25cm} \rho {\textstyle{\sum_{m=0}^{2}}}{\left\| {{{\bf{x}}_m} - {\rm{ }}{\bm{\bar \theta }}^{(n+1)} + {\bm{\mu }}_m^{(n)}} \right\|^2} +{\rm{tr}}({{\bf{S}}^{(n+1)}})u   }_{
			\displaystyle{\hspace{0.5cm}- \eta {\rm{tr}}\left( {\left( {{\bf{I}} - {{\bf{S}}^{(n+1)}}} \right){\bf{Gw}}{{\bf{w}}^H}{{\bf{G}}^H}\left( {{\bf{I}} - {{\bf{S}}^{(n+1)}}} \right)} \right)   }
	}}\\
	&s.t. \hspace{0.1cm}{\rm{C}}8,\,{\rm{C}}9, \, {\rm{C}}10.
\end{split}
\end{equation} 
In problem ({\bf{P2.5}}),  ${{\bf{x}}_0}$, ${{\bf{x}}_1}$ and ${{\bf{x}}_2}$ are independent from each other. Therefore, problem ({\bf{P2.5}}) can be decomposed into subproblems.The subproblem for obtaining ${{\bf{x}}_0^{(n+1)}}$ can be given as
\begin{equation}
\begin{split}
\nonumber
({\bf{P2.6}})\hspace{-0.1cm}:\{{{\bf{x}}_0^{(n+1)}}\} =\mathop {{\rm{argmin}} }\limits_{{\bf{x}}_0} \rho {\left\| {{{\bf{x}}_0} - {\rm{ }}{\bm{\bar \theta }}^{(n+1)} + {\bm{\mu }}_0^{(n)}} \right\|^2} s.t. {\rm{C}}10.
\end{split}
\end{equation}  
Problem ({\bf{P2.6}}) is non-convex but has only one constraint. We can derive ${{\bf{x}}_0^{(n+1)}}$ in closed form as follows
\begin{equation}
\label{Obtx0}
{{\bf{x}}_0^{(n+1)}} = \left\{ {\begin{array}{*{20}{l}}
	{{{\bf{x}}_{0,[1]}^{(n+1)}}, \,\, \text{if}\,\,\alpha L{\left| {{\bf{b}}_{}^H\left( {{\bm{\bar \theta }}_{}^{(n+1)} - {\bm{\mu }}_{0}^{(n)}} \right)} \right|^2}/{{\sigma ^2}} \ge {\gamma _B}}\\
	{{{\bf{x}}_{0,[2]}^{(n+1)}}, \,\, \text{else if} \,\,\,\, {\bf{b}}_{}^H\left( {{\bm{\bar \theta }}_{}^{(n+1)} - {\bm{\mu }}_{0}^{(n )}} \right) = 0}\\
	{{{\bf{x}}_{0,[3]}^{(n+1)}}, \,\, \text{otherwise}}
	\end{array}} \right.
\end{equation}
where ${{\bf{x}}_{0,[1]}^{(n+1)}} = {\bm{\bar \theta }}_{}^{(n+1)} - {\bm{\mu }}_{0}^{(n )}$, ${{\bf{x}}_{0,[2]}^{(n+1)}} = {{\bf{x}}_{0,[1]}^{(n+1)}} + \sqrt {\frac{{{\sigma ^2}{\gamma _B}}}{{\alpha L}}} {\rm{ }}{\bf{b}}/{{{{\left\| {\bf{b}} \right\|}^2}}}$ and ${{\bf{x}}_{0,[3]}^{(n+1)}} = {{\bf{x}}_{0,[1]}^{(n+1)}} + \big(\sqrt {\frac{{{\sigma ^2}{\gamma _B}}}{{\alpha L}}}  - \big| {{\bf{b}}_{}^H{{\bf{x}}_{0,[1]}^{(n+1)}}} \big|\big){}{\bf{bb}}_{}^H{{\bf{x}}_{0,[1]}^{(n+1)}}\big/\big({{{\left\| {\bf{b}} \right\|}^2}\big| {{\bf{b}}_{}^H{{\bf{x}}_{0,[1]}^{(n+1)}}} \big|}\big)$.
The derivation can be seen in Appendix. Trough the similar approach, the closed form expressions of ${{\bf{x}}_1^{(n+1)}}$ and ${{\bf{x}}_2^{(n+1)}}$ can be obtained and are omitted here because of the page limit.

(c) Dual Variables Update

At the end of the $(n+1)$-th iteration of the ADMM procedure, the dual variables can be updated as
\begin{equation}
\label{Obtmu}
{{\bm{\mu }}_m^{(n+1)}} = {\bm{\mu }}_{m}^{(n)} + {\bf{x}}_{m}^{(n+1)} - {{\bm{\bar \theta }}^{(n+1)}}, m=0,1,2.
\end{equation}

The ADMM procedure discussed above yields a suboptimal solution to problem  ({\bf{P2.3}}) denoted by $\{{\bf{S}}^+,{\rm{ }}{\bm{\bar \theta }}^+,{\bf{x}}_0^+,{\bf{x}}_1^+,{\rm{ }}\bf{x}_2^+\}$. Let $\bm{\Psi}^+=diag((\bar{\theta}^+_1/\bar{\theta}^+_{I_R+1})^\dag,...,(\bar{\theta}^+_{I_R}/\bar{\theta}^+_{I_R+1})^\dag)$. Then, from the mapping relation between  ({\bf{P2.3}}) and  ({\bf{P2.1}}), we know $\{{\bf{S}}^+,\bm{\Psi}^+\}$ is a suboptimal solution to  ({\bf{P2.1}}).

\vspace{0.2cm}
\noindent\textit{C. Overall Algorithm}

According to the BCD technique, by alternatively optimizing the transmit beamforming vector ${\bf{w}}$, the RIS element mode assignment diagonal matrix  ${\bf{S}}$ and the reflect beamforming diagonal matrix ${\bm{\Psi}}$ based on Sections III-A and III-B, a suboptimal solution to the original transmission design problem ({\bf{P}}) can be found. The algorithm is summarized in Algorithm 2.  {\textcolor{black}{If $E_{\text{RIS}}({{\bf{w}}^{(k+1)},{{\bf{S}}}^{(k)}}) \ge E_{\text{RIS}}({{\bf{w}}^{(k+1)},{{\bf{S}}}^{(k+1)}})$ holds (which is always the case), the objective value of ({\bf{P}}) can keep nonincreasing over the iterations. Therefore, the convergence of Algorithm 2 is guaranteed.  }}

 {\textcolor{black}{The complexity of Algorithm 2 mainly comes from solving problem ({\bf{P1.2}}) and updating ${\bf{x}}_m$'s in each iteration of the ADMM procedure. The complexity of solving the convex SDP problem ({\bf{P1.2}}) is $\mathcal{O}(N^{7})$ \cite{Han41} and the complexity of updating ${\bf{x}}_m$ is $\mathcal{O}(I_R^2)$. Therefore, the complexity of Algorithm 2 is $\mathcal{O}((N^{7} + I^2_RI_{AD}){I}_{BD})$ where $I_{AD}$ and ${I}_{BD}$ represent the numbers of the ADMM and the BCD iterations in Algorithm 2 respectively. }}
\begin{algorithm}[h] \label{AL2}
	\caption{Transmission Design Algorithm for RIS Power Consumption Minimization}
{\small	\begin{algorithmic}[1]
		\REQUIRE $\bf{h}_d$,\, $\bf{h}_r$,\,$\bf{G}$,\,$P_{bgt}$,\,$\sigma_2$,\,$\alpha$,\,$\mu$,\,$\eta$,\,$\gamma_A$,\,$\gamma_B$
		
		\ENSURE Solution  $\{\bf{w}^{(\bar k)},{\bf{S}}^{(\bar k)},\bm{\Psi}^{(\bar k)}\}$ and the achieved minimum RIS power consumption $E^{min}_{RIS}= - \eta {\rm{tr}}( ( {{\bf{I}} - {\bf{S}}^{(\bar k)}} ){\bf{Gw}}^{(\bar k)}$ ${({{\bf{w}}^{(\bar k)}})^H}{{\bf{G}}^H}( {{\bf{I}} - {\bf{S}}^{(\bar k)}} ) )$ $+{\rm{tr}}({\bf{S}}^{(\bar k)})u$ where ${\bar{k}}$ is the iteration index at which the BCD iterations stop.
		
		\STATE Initialize ${\bf{S}}=\bf{I}$, ${\bm{\Psi}}=\bf{I}$ and the index for BCD iterations $k=0$.
		
		\REPEAT
		
		\STATE Obtain an optimal solution ${\bf{w}}^+$ to ({\bf{P1.1}}) by solving problem ({\bf{P1.2}}) and applying Algorithm 1, set ${\bf{w}}^{(k+1)}={\bf{w}}^+$ and fix ${\bf{w}}={\bf{w}}^{(k+1)}$.
		
		\STATE Initialize the index for ADMM iterations $n=0$. If $k=0$, initialize ${\bf{x}}_m^{(0)}={\bf{1}}, {\bm{\mu}}_m^{(0)}={\bf{0}}, m=0,1,2$, otherwise, initialize ${\bf{x}}_m^{(0)}={\bf{x}}_m^+, {\bm{\mu}}_m^{(0)}={\bm{\mu}}_m^+, m=0,1,2$.
		
		\REPEAT
				
		\STATE  Update ${\bf{S}}^{(n+1)}$ and ${\bm{\bar \theta }}^{(n+1)}$ through \eqref{Obthetai}-\eqref{Obtsi}.
		
		\STATE  Update ${\bf{x}}_m^{(n+1)},m=0,1,2$ through \eqref{Obtx0} and the corresponding closed form expressions.
		
		\STATE  Update ${\bm{\mu}}_m^{(n+1)},m=0,1,2$ through \eqref{Obtmu}.
		
		\STATE Update n=n+1.
		
		\UNTIL{Convergence}
		
		\STATE Set ${\bf{S}}^+={\bf{S}}^{(\bar{n})}$,${\bm{\bar \theta }}^+={\bm{\bar \theta }}^{(\bar{n})}$,${\bf{x}}_m^+={\bf{x}}_m^{(\bar n)}$,${\bm{\mu}}_m^+={\bm{\mu}}_m^{(\bar n)}$,$m=0,1,2$ and $\bm{\Psi}^+=diag((\bar{\theta}^+_1/\bar{\theta}^+_{I_R+1})^\dag,...,$ $(\bar{\theta}^+_{I_R}/\bar{\theta}^+_{I_R+1})^\dag)$ where ${\bar{n}}$ is the iteration index at which the ADMM procedure converges.
		
		\STATE Set ${\bf{S}}^{(k+1)}={\bf{S}}^+$ and $\bm{\Psi}^{(k+1)}=\bm{\Psi}^+$ and fix  ${\bf{S}}={\bf{S}}^{(k+1)}$ and $\bm{\Psi}=\bm{\Psi}^{(k+1)}$.
		
		\STATE Update k=k+1.
		
		\UNTIL{the objective value of the original  transmission design problem ({\bf{P}}) stop decreasing.} 
		
	\end{algorithmic}}  
\end{algorithm}
\vspace{-0.2cm}
\section{Simulation Results}
\label{sect:numerical}

In this section, we demonstrate numerical results on the performance of the proposed transmission design algorithm (which is referred to
as P-A).  {\textcolor{black}{For all the examples, the A-Tx-RIS, RIS-C-Rx and A-Tx-C-Rx distances are given by $15$, $30$ and $40$ in meters. The A-Tx-B-Tx and B-Tx-C-Rx channels follow the Rician fading channel model and the Rician factor is set to 3. As for  A-Tx-C-Rx channels, we consider the Rayleigh fading case.}} We set $\alpha=1$, $\eta=1$ and {\textcolor{black}{$L=50$}} and $u=15$uW \cite{ZouS}. The transmit power budget is set as {\textcolor{black}{$30$dBm}} and the noise power is $-45$dBm.

{\textcolor{black}{For comparison, we also consider a benchmark scheme (which is referred to as B-S) in simulations. In B-S, the BCD technique is also applied for tackling the original transmission design problem ({\bf{P}}) and the transmit beamforming is optimized based on the discussion in Section III.A; but, the RIS element mode assignment is optimized through the successive convex approximation (SCA) approach as shown in \cite{Sun2018}, and the reflect beamforming is optimized by SDR. The complexity of B-S can be given by $\mathcal{O}((N^{7} + I^7_RI_{SC})\bar{I}_{BD})$  where $I_{SC}$ and $\bar{I}_{BD}$ denote the numbers of the SCA and the BCD iterations in B-S. It can be seen that, compared with B-S, P-A  has an advantage in complexity for large $I_R$ cases. }}

\begin{figure}
	\begin{center}
		\includegraphics[width=0.7\linewidth, draft=false]{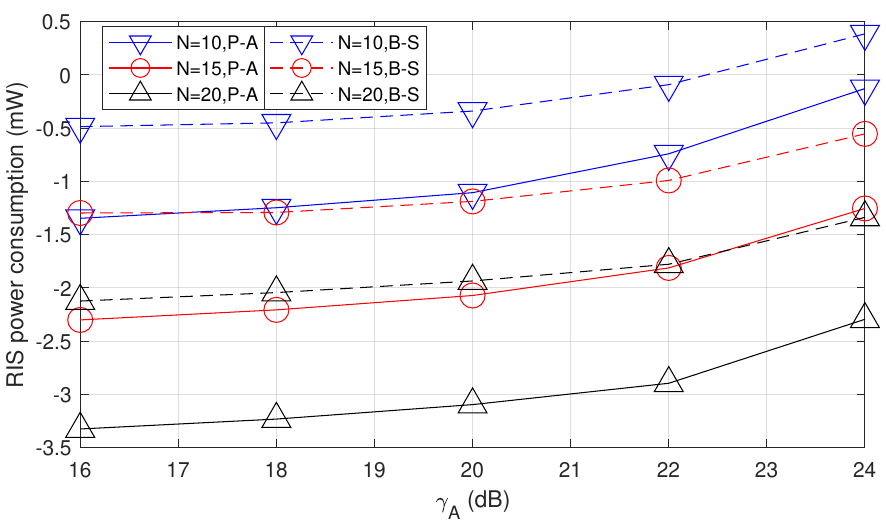}
		\caption{ The minimum RIS power consumption versus $\gamma_A$ with $I_R=100$, $\gamma_B=10$dB and different values of $N$.}
	\end{center}
\end{figure}

{\textcolor{black}{Fig. 2 represents the minimum RIS power consumption achieved by P-A  and B-S versus $\gamma_A$ with $I_R=100$, $\gamma_B=10$dB and different values of $N$.  
 It can be observed that P-A performs much better than its counterpart.  Moreover, with more antennas equipped on the A-Tx, P-A can achieve better performance. }} {\textcolor{black}{In Fig.3, the minimum RIS power consumption achieved by P-A and B-S are plotted against increasing $\gamma_B$ with $N=10$, $\gamma_A=15$dB and different values of $I_R$. As can be seen from Fig. 2 and Fig. 3, in most cases,  P-A can keep RIS power consumption less than 0. This implies that, through the proposed algorithm, the RIS with EH can provide energy compensation to the B-Tx in stead of consuming its power in quite some cases.}}

\begin{figure}
	\begin{center}
		\includegraphics[width=0.7\linewidth, draft=false]{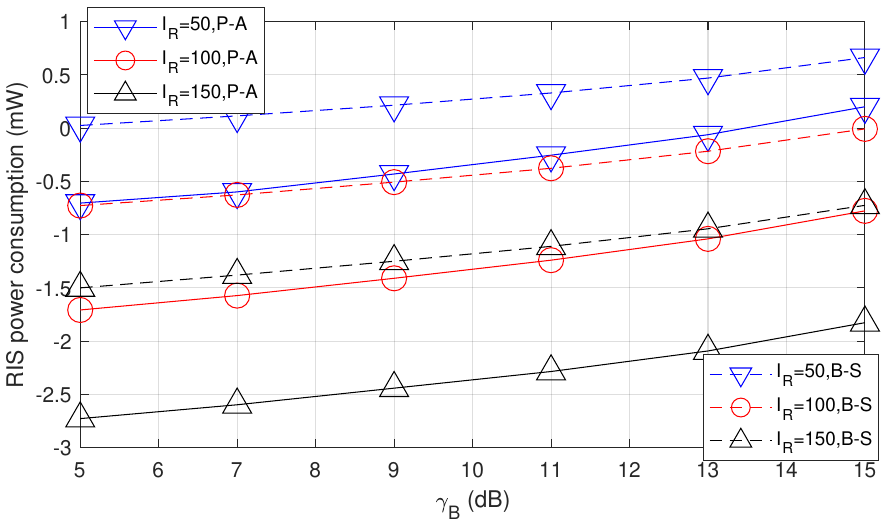}
		\caption{ The minimum RIS power consumption versus $\gamma_B$ with  $N=10$, $\gamma_A=15$dB and different values of $I_R$.}
	\end{center}
\end{figure}

\vspace{-0.1cm}
\section{Conclusions}
\label{sect:conclusion}

In this letter, we considered a RIS assisted CABC system where each element on the RIS can operate under either signal reflecting mode or EH mode and investigated a transmission design problem to minimize the RIS power consumption with QoS constraints. To tackle the optimization problem, we developed an algorithm by employing the BCD, SDR and ADMM techniques. Simulation results verified the effectiveness of the proposed algorithm.

\vspace{-0.1cm}
\section{Appendix}

From Problem ({\bf{P2.6}}), we can observe that if ${\bm{\bar \theta }}^{(n+1)} - {\bm{\mu }}_0^{(n)}$ is feasible, it must be an optimal solution. Thus, we have ${{\bf{x}}_{0}^{(n+1)}} = {{\bf{x}}_{0,[1]}^{(n+1)}}$, if $\alpha L{| {{\bf{b}}_{}^H( {{\bm{\bar \theta }}_{}^{(n+1)} - {\bm{\mu }}_{0}^{(n)}} )} |^2}/{{\sigma ^2}} \ge {\gamma _B}$. 

If ${\bm{\bar \theta }}^{(n+1)} - {\bm{\mu }}_0^{(n)}$ does not satisfy constraint C10, optimal solutions to Problem ({\bf{P2.6}}) must be achieved when the equality in constraint C10 holds. According to the discussion in \cite[Section III.A]{Huang2016}, we have ${{\bf{x}}_0^{(n+1)}}$ must take the form 
\begin{equation}
\begin{split}
\label{Appen}
{{\bf{x}}_0^{(n+1)}}=& ( {{\bm{\bar \theta }}_{}^{(n + 1)} - {\bm{\mu }}_{0}^{(n)}} ) +\\
& \frac{{\sqrt {\frac{{{\sigma ^2}{\gamma _B}}}{{\alpha L}}} {e^{j\varphi }} - {\bf{b}}_{}^H( {{\bm{\bar \theta }}_{}^{(n + 1)} - {\bm{\mu }}_{0}^{(n)}} )}}{{{{\| {\bf{b}} \|}^2}}}{\bf{b}}.
\end{split}
\end{equation}
where $\varphi$ is an angle that can minimize the objective value of Problem ({\bf{P2.6}}). If ${\bf{b}}_{}^H( {{\bm{\bar \theta }}_{}^{(n + 1)} - {\bm{\mu }}_{0}^{(n)}})=0$, we have ${{\bf{x}}_0^{(n+1)}}={{\bf{x}}_{0,[2]}^{(n+1)}}$. Otherwise, by choosing $\varphi$ to be the angle of ${\bf{b}}_{}^H( {{\bm{\bar \theta }}_{}^{(n + 1)} - {\bm{\mu }}_{0}^{(n)}} )$, the minimum objective value of ({\bf{P2.6}}) can be achieved and we can get ${{\bf{x}}_0^{(n+1)}}={{\bf{x}}_{0,[3]}^{(n+1)}}$.

\vspace{-0.1cm}



\ifCLASSOPTIONcaptionsoff
  \newpage
\fi

\bibliographystyle{IEEEtran}
\bibliography{IEEEabrv,refabrv}

\end{document}